\documentclass[aip,rsi,amsmath,amssymb,preprint]{revtex4-1} 
\usepackage{graphicx}
\usepackage{dcolumn}
\usepackage{gensymb}
\usepackage[noabbrev]{cleveref}
\usepackage[mathlines]{lineno}

\begin{document}


\title{Novel circuit design for high-impedance and non-local electrical measurements of two-dimensional materials}

\author{Adolfo De Sanctis}
\affiliation{Centre for Graphene Science, College of Engineering, Mathematics and Physical Sciences, University of Exeter, Exeter EX4 4QF, United Kingdom}
\author{Jake D. Mehew}
\affiliation{Centre for Graphene Science, College of Engineering, Mathematics and Physical Sciences, University of Exeter, Exeter EX4 4QF, United Kingdom}
\author{Saad Alkhalifa}
\affiliation{Centre for Graphene Science, College of Engineering, Mathematics and Physical Sciences, University of Exeter, Exeter EX4 4QF, United Kingdom}
\affiliation{Department of Physics, College of Science, University of Duhok, Duhok, Kurdistan Region, Iraq}
\author{Callum P. Tate}
\affiliation{Centre for Graphene Science, College of Engineering, Mathematics and Physical Sciences, University of Exeter, Exeter EX4 4QF, United Kingdom}
\author{Ashley White}
\affiliation{Centre for Graphene Science, College of Engineering, Mathematics and Physical Sciences, University of Exeter, Exeter EX4 4QF, United Kingdom}
\author{Adam R. Woodgate} 
\affiliation{Centre for Graphene Science, College of Engineering, Mathematics and Physical Sciences, University of Exeter, Exeter EX4 4QF, United Kingdom}
\author{Monica F. Craciun}
\affiliation{Centre for Graphene Science, College of Engineering, Mathematics and Physical Sciences, University of Exeter, Exeter EX4 4QF, United Kingdom}
\author{Saverio Russo}
\email[Correspondence should be addressed to: ]{S.Russo@exeter.ac.uk}
\affiliation{Centre for Graphene Science, College of Engineering, Mathematics and Physical Sciences, University of Exeter, Exeter EX4 4QF, United Kingdom}

\begin{abstract}
\textbf{ABSTRACT.} 
Two-dimensional materials offer a novel platform for the development of future quantum technologies. However, the electrical characterisation of topological insulating states, non-local resistance and bandgap tuning in atomically-thin materials, can be strongly affected by spurious signals arising from the measuring electronics. Common-mode voltages, dielectric leakage in the coaxial cables and the limited input impedance of alternate-current amplifiers can mask the true nature of such high-impedance states. Here, we present an optical isolator circuit which grants access to such states by electrically decoupling the current-injection from the voltage-sensing circuitry. We benchmark our apparatus against two state-of-the-art measurements: the non-local resistance of a graphene Hall bar and the transfer characteristic of a WS$_2$ field-effect transistor. Our system allows the quick characterisation of novel insulating states in two-dimensional materials with potential applications in future quantum technologies.
\end{abstract}


\maketitle 

A widely and long-accepted classification of the materials and states of matter relies on the electrical conductivity of systems which are metallic, insulating or semiconducting. However, their characterization has often proven to be very challenging. Indeed, the difficulty of measuring a superconducting zero-state resistance faced in the early days by Kamerling Onnes\cite{KamerlinghOnnes1911} is nowadays being emulated by the difficulty of measuring highly resistive states under extreme conditions (e.g. high magnetic field and low temperature), in systems with extremely low dimensionality such as atomically-thin materials\cite{Geim2007,Wang2012,Chhowalla2013}. The recent observations of insulating states of topological origin\cite{Kosterlitz2016} or induced by many-body interactions\cite{Qi2011}, a non-saturating linear magneto-resistance of quantum mechanical origin\cite{Yuasa2004}, the observation of giant non-locality\cite{Abanin2011}, viscous electron flow\cite{Bandurinaad2016} and tunable bandgap\cite{Craciun2009} in single- and few-layer graphene devices are examples of recently discovered high-impedance (Hi-Z) states and phenomena of high interest for their potential applications in the field of quantum technologies. However, contrasting experimental reports in high quality devices and materials suggest that the electronics used in the characterization of these states needs to be reconsidered to eliminate spurious signals.

There is a wide consensus that alternate-current (AC) lock-in measurements offer higher resolution and greater noise rejection than direct-current (DC) techniques and, for this reason, they are widely used to characterize the response of electronic devices as well as in many practical applications, including medical diagnostics\cite{Goovaerts1998}, classical and quantum metrology\cite{Kotler2011}. Careful consideration needs to be given to the characterization of insulating states with this technique. For example, a spurious negative value of resistance can appear in the non-local transport of a graphene Hall bar\cite{Sui2015,Shimazaki2015} due to the incomplete rejection of the common-mode voltage (CMV) present at the input of lock-in amplifiers. Such spurious voltage is the direct consequence of the finite input impedance of the measuring amplifier and the dielectric leakage of the transmission lines, usually BNC-terminated coaxial cables. DC amplifiers and sources might offer a straightforward solution since they typically have a higher input impedance than their AC counterparts. However, they are characterized by a larger noise floor than AC electronics making the characterization of insulating states with narrow energy gaps, such as in 2D materials\cite{Craciun2009,Overweg2017}, challenging.

In this letter, we present an experimental apparatus designed specifically to overcome the aforementioned limitations, allowing us to eliminate spurious artefacts and grant direct access to the aforementioned Hi-Z states. The key to achieving this goal lies in the ability to decouple electrically (float) the circuits connected to the current-injection probes from that of the voltage-sensing contacts. To this end, we developed a battery-powered optical isolator (or optocoupler) circuit, together with a custom-designed measuring chamber which is able to float both core and shell of the coaxial cables used to interface the device to the instruments. In this manuscript we present two versions of this circuit: low-power and low-noise. The low-noise circuit is suitable for high-precision measurements as it is characterised by a noise level well below the intrinsic noise of the commercial measuring instrument used in these experiments ($<10^{-5}\,\mathrm{V/\sqrt{Hz}}$ at frequencies below $100\,\mathrm{Hz}$). On the other hand, the low-power circuit demonstrates an increase in the period between battery charges from $4$ to more than $30$ days with a noise floor of $\sim10^{-4}\,\mathrm{V/\sqrt{Hz}}$, making it ideal for remote, long-term, operation. We benchmark the use of these circuits in two state-of-the-art experiments: measurement of the non-local resistance in a graphene Hall bar in perpendicular magnetic field\cite{Abanin2011} and AC electrical characterisation of an atomically thin WS$_2$ field-effect transistor. Finally, the reduced number of components and the possibility of miniaturisation using surface-mount-devices makes our approach of interest for applications with stringent requirements on integration such as particle accelerators and space applications.

\begin{figure*}{}
	\includegraphics[scale=1]{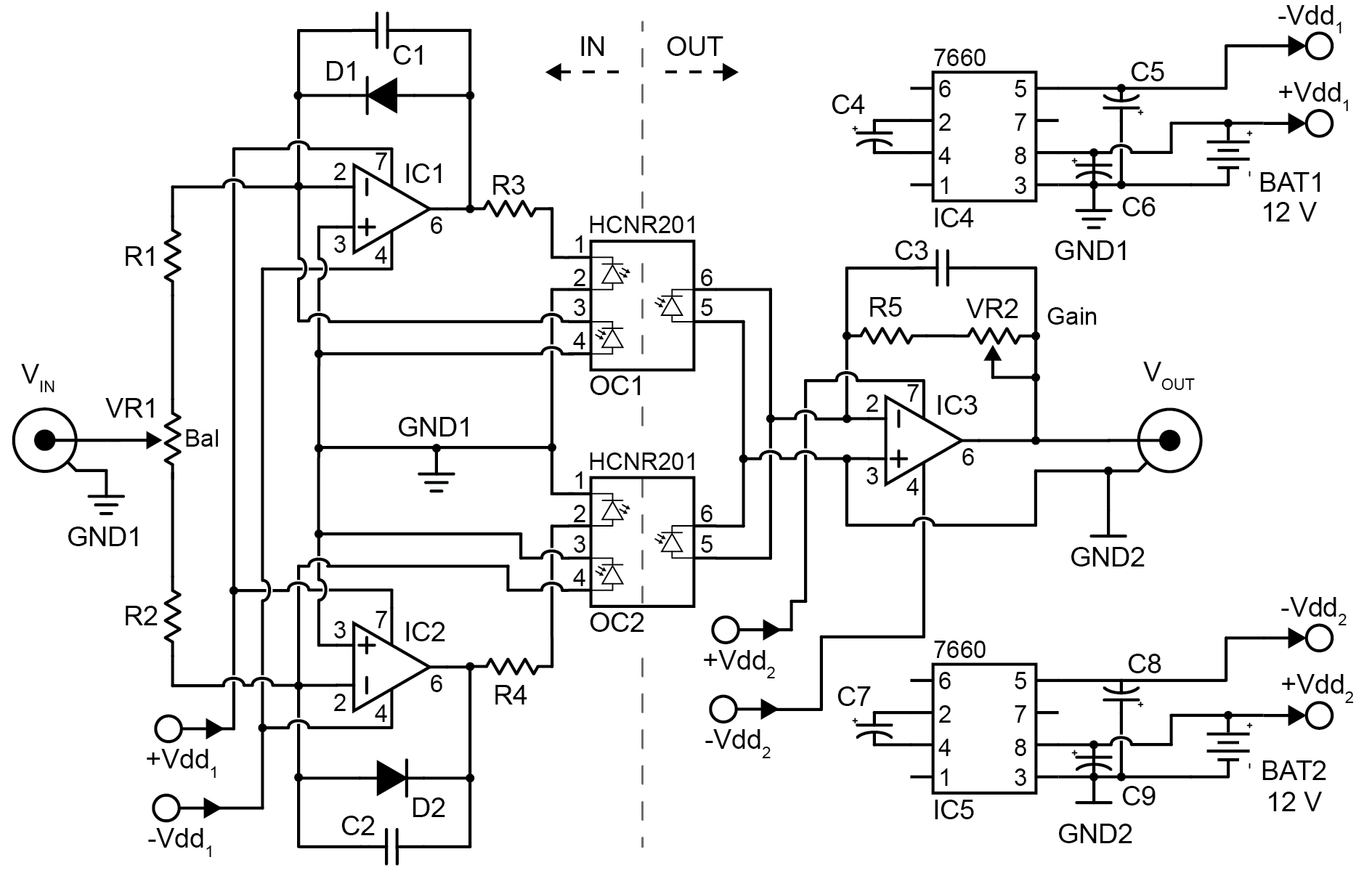}
	\caption{\label{Figure1}\textbf{Electronic schematic of the bipolar optocoupler circuit.} The circuit is designed around two HCNR201 high-linearity optocouplers (OC1, OC2) powered by two $12\,\mathrm{V}$ batteries. Two ICL7660 CMOS voltage converters (IC4 and IC5) are used to supply the required dual voltage ($\pm12\,\mathrm{V}$) to the operational amplifiers (IC1, IC2 and IC3). The vertical dashed line indicates the optical coupling point between the input (IN) and the output (OUT) stages of the circuit.}
\end{figure*}

\section{Operating principles and instrumentation} \label{sec:OperPrinciples}
To allow the measurement of Hi-Z devices in AC and prevent common-mode voltage (CMV), inductive coupling, and dielectric leakage currents from affecting the measuring instruments, it is necessary to effectively float the voltage probes of such instruments from the current-injection leads of the device under test. To achieve this goal, we incorporate into our apparatus an optical isolator or optocoupler. This is a device which converts an electric signal into light using a light-emitting-diode (LED) and converts it back into an electrical signal through one, or more, photodiodes. An ideal optocoupler has a linear response with a gain of $1$. Whilst optical coupling is a well-known technique to isolate electric signals, crucially in the specific case of AC lock-in measurements, the optical isolation of the core signal of the wire is not sufficient to eliminate the emergence of spurious electrical signals. Indeed, the dielectric leakage in the coaxial cables (including connectors and terminations) between the core wire and the ground shielding causes the signal in one section of the circuit to be coupled to the ground, known as ground potential difference\cite{Morrison_GND}, and therefore to be measured elsewhere in the circuit. This coupling is one kind of CMV present at the input of a differential amplifier. For this reason, in AC measurements, it is also necessary to decouple the elements of the circuit used in defining the electrical ground. We realise such decoupling by using a carefully designed measuring probe which is able to isolate the cores from the grounding shells of the BNC wires with a resistance in excess of $1\mathrm{T\Omega}$.

\begin{figure*}{}
	\includegraphics[scale=1]{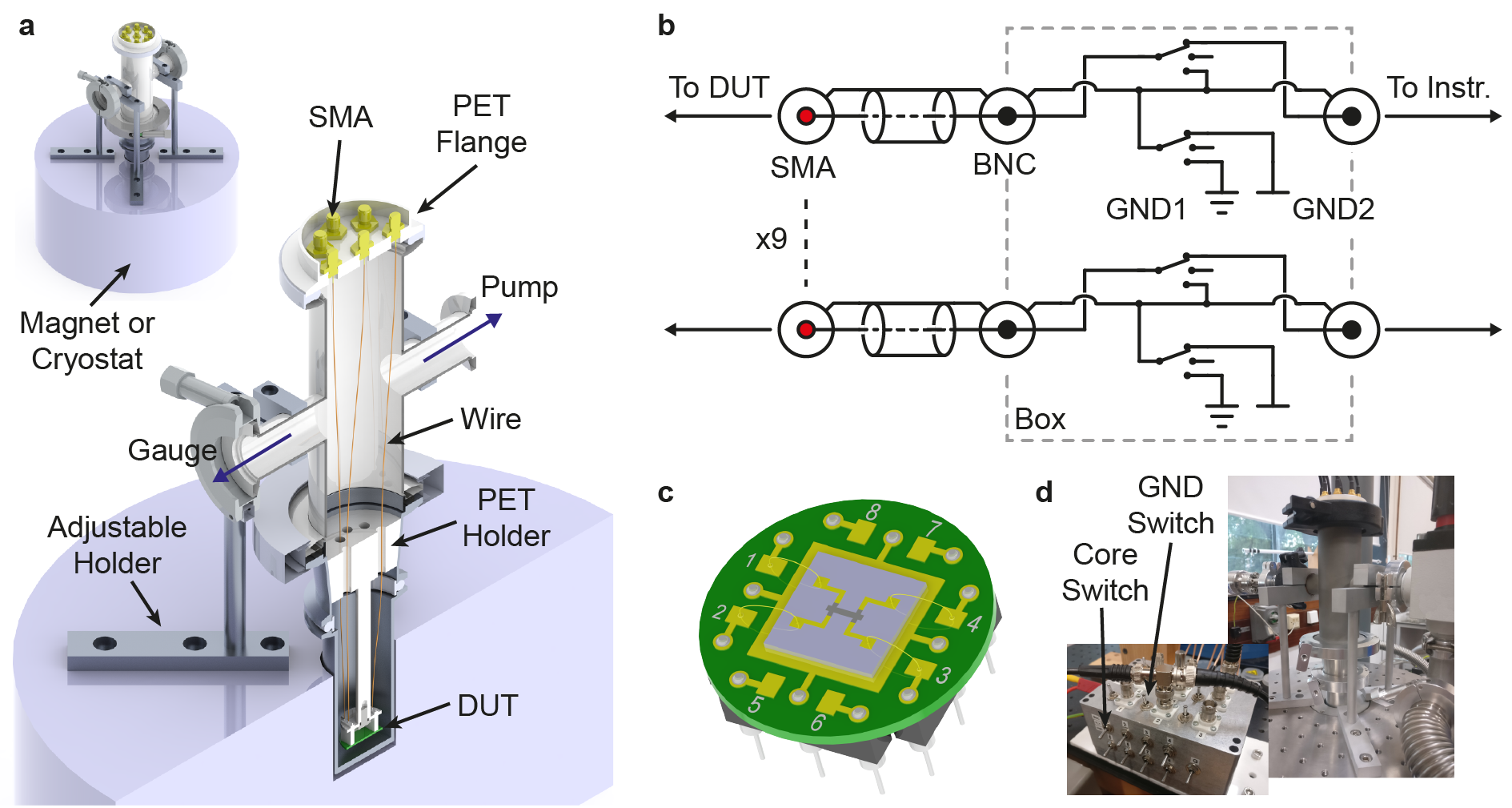}
	\caption{\label{Figure2}\textbf{High-impedance measuring chamber and circuitry.} \textbf{a}, Schematic drawing of the vacuum chamber used to perform the measurements, configured to retrofit a magnet or cryostat. In order to avoid stray capacitance, each SMA connector is wired to the device under test (DUT) using a single-core insulated copper wire. \textbf{b}, Schematic of the break-out box used to interface the DUT to the measuring instruments. For each BNC connector it is possible to choose two separate terminals for the shielding (GND1 and GND2), in order to ensure complete floating of the measuring probes. \textbf{c}, Model of the printed circuit board (PCB) used to mount the DUT. The pins are spaced $5$ mm apart to ensure high insulation. \textbf{d}, Photograph of the actual chamber mounted on an \textit{Oxford Instruments} Microstat MO superconducting magnet.}
\end{figure*}

The core electronic circuit is shown in \Cref{Figure1}. This consists of a bipolar optocoupler with integrated dual supply based on two HCNR201 high-linearity analog optocoupler (OC1 and OC2) integrated circuits (ICs). Such ICs comprise a LED and two photodiodes. A primary photodiode is used to transfer the input signal to the output while the secondary photodiode is used as a feedback to ensure optimal linearity and stability. This feature makes the performance of the HCNR201 very close to that of an ideal optocoupler. Indeed, the typical nonlinearity (i.e. the end-point deviation from a straight line in the relationship between the current measured from the first photodiode versus the current measured from the second, $I_{pd}$) of the device is $0.01\,\%$, with a maximum of $0.05\,\%$ under test conditions of $5\,\mathrm{nA}<I_{pd}<50\,\mathrm{mA}$ and $0\,\mathrm{V}<V_{pd}<15\,\mathrm{V}$, while the temperature coefficient of the HCNR201 is $-0.3\,\%/\degree \mathrm{C}$ in the range $-40\,\degree \mathrm{C}$ to $85\,\degree \mathrm{C}$, corresponding to a maximum nonlinearity of $0.07\,\%$ in this range\cite{HCNR_Datasheet}. The input signal (V$_\mathrm{IN}$) is fed through a balance variable resistor (VR1) to the LED stage of the optocouplers. Each of the secondary photodiodes is wired in a feedback loop comprising an operational-amplifier (Op-Amp, IC1 and IC2, respectively) and a diode (D1 and D2, respectively). The diodes are wired such that the positive and negative parts of the signal are transmitted through one or the other HCNR201. The balance resistor VR1 is used to compensate for the resistance mismatch in the input stage between the two paths of the circuit, and it is adjusted before operation in order to obtain the same voltage swing in the positive and negative part of the AC signal. The shell of the input BNC is ground in common with the input stage of the HCNR201. The output stage is composed by the HCNR201 primary photodiodes and an operational amplifier (IC3) with a gain of $1$ (regulated through VR2). The grounding route of the output stage is isolated from the one of the input. The power-supply circuitry comprises two $12\,\mathrm{V}$ lithium-polymer (LiPo) batteries (one for the input and one for the output stage) and two ICL7660A  voltage inverters (IC4 and IC5), which generate the dual voltage ($\pm12\,\mathrm{V}$) necessary for the correct operation of the Op-Amps. The two voltage inverters avoid the use of two pairs of batteries to generate a dual voltage, reducing the size of the instrument. Finally, in the low-power design we use three LT1097 Op-Amps for the input and output stages (IC1, IC2 and IC3), whilst in the low-noise one, these are replaced by three LT1028 Op-Amps. The full list of components is reported in table S1 in Supplementary Information.

The isolation of the electrical connections used to wire the measuring instruments (such as lock-in and voltage amplifiers, oscilloscopes and spectrum analysers) to the device under test is accomplished using the design shown in \cref{Figure2}. A vacuum chamber is equipped with a polyethylene terephthalate (PET) vacuum flange onto which $9$ vacuum-compatible Sub-miniature version A (SMA) connectors are mounted (see \cref{Figure2}a and figure S2c in Supplementary Information). The PET construction of the flange ensure high insulation between the shells of each of the SMA connectors ($>1\,\mathrm{T\Omega}$). The central pin of each connector is wired directly to the sample holder with a single-core copper wire. The device under test is mounted on a custom-built printed-circuit-board in which high insulation between each contact pin is ensured by doubling the spacing usually used in single-in-line commercial connectors, from $2.54\,\mathrm{mm}$ to $5.08\,\mathrm{mm}$, as shown in \cref{Figure2}c. The most important part of the apparatus is the break-out box used to connect each SMA connector to the measuring instrumentation, which has to maintain the high level of electrical insulation obtained in the vacuum chamber. This is shown schematically in \cref{Figure2}b. Crucially, the use of mechanical switches allow us to connect each core and each shell independently to one ground line (GND1) or another (GND2). A picture of the actual apparatus is shown in \cref{Figure2}d. The whole assembly is designed to retrofit into a number of measurement set-ups including cryostats\cite{Khodkov2015}, dipping probes, and optical microscopes\cite{DeSanctis2017_G16} (for our purposes this has been fitted to an \textit{Oxford Instruments} Microstat MO superconducting magnet).

\begin{figure}{}
	\includegraphics[scale=1]{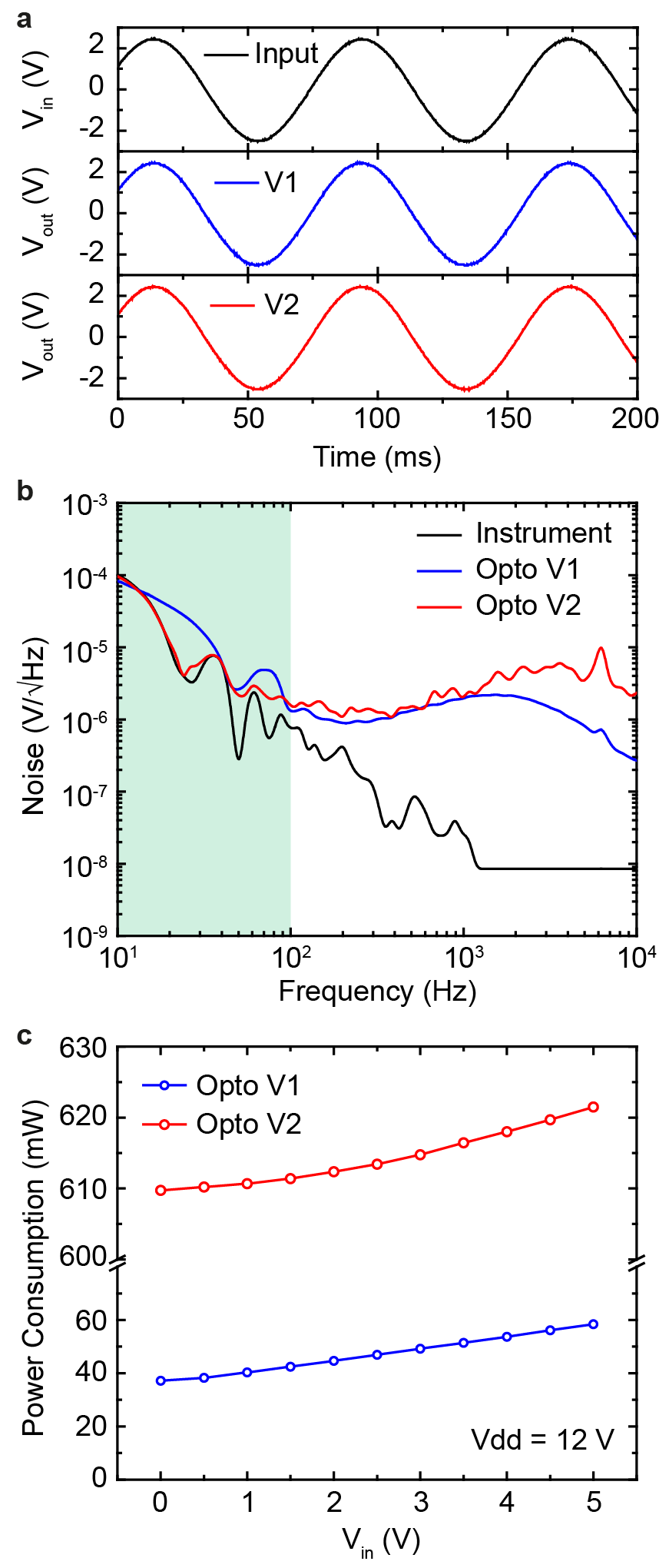}
	\caption{\label{Figure3}\textbf{Performance of the optocoupler circuits.} \textbf{a}, Input versus output signal for the low-power (V1) and low-noise (V2) versions. \textbf{b}, Noise as a function of frequency for the two versions compared to the intrinsic noise of the measuring lock-in amplifier. The low-frequency region is highlighted in green. \textbf{c}, Power consumption as a function of input voltage (peak-to-peak) for the two versions of the optocoupler circuit.}
\end{figure}

The circuit shown in \cref{Figure1} is assembled on a printed circuit board and mounted in a metal box for shielding, as shown in figure S2a and figure S2b in Supplementary Information, using through-hole components. This circuit can be easily adapted to use surface-mount components for applications in which space is a premium. The fidelity, i.e. the ability to reproduce the input signal (V$_\mathrm{in}$) at the output (V$_\mathrm{out}$), of the two versions of the circuit is shown in \cref{Figure3}a (where V1 refers to the low-power and V2 to the low-noise version). No substantial difference is observed in the output curves from the input ones for both versions. 

The noise as a function of frequency has been measured using an \textit{Ametek} Model 7270 DSP Lock-in amplifier in Noise-measurement mode. \Cref{Figure3}b shows the noise from the lock-in (black line) and the two versions of the optocoupler. We can see that the low-power version has a noise level of $>10^{-5}\,\mathrm{V/\sqrt{Hz}}$ with a pronounced peak around $100\,\mathrm{Hz}$. At higher frequencies the noise level is between one and two orders of magnitude higher than the lock-in intrinsic level ($<10^{-7}\,\mathrm{V/\sqrt{Hz}}$), which is mostly due to the noise performance of the Op-Amps used. The sharp peak observed at $6\,\mathrm{kHz}$ corresponds to the frequency at which the ICL7660A operates to produce the negative power supply voltage. On the contrary, the low-noise version shows a noise level at low frequency identical to the lock-in amplifier ($<10^{-5}\,\mathrm{V/\sqrt{Hz}}$), indicating that its intrinsic noise level is at least one order of magnitude smaller than that of the lock-in. The same behaviour of the low-power version is observed at higher frequencies, albeit with slightly higher noise level above $1\,\mathrm{kHz}$. In both versions, the frequency of the ICL7660A is easily adjustable in a wide range by adding a simple capacitor to the circuit (between pin $7$ and GND in \cref{Figure1}), allowing for the customisation of the high-frequency noise level of the optocoupler.

\Cref{Figure3}c shows the power consumption as a function of input voltage for each model. We can see that the low-noise version uses roughly ten times as much power as the low-power one. This is due to the higher current drained by the LT1028 low-noise Op-Amp, since the input transistors of this Op-Amp are operated at nearly $1\,\mathrm{mA}$ of collector currents to achieve low voltage noise (since voltage noise is inversely proportional to the square root of the collector current)\cite{Asparuhova2004}. For the purpose of remote application of our circuit we estimate that a $4\,\mathrm{Ah}$ battery (the one used in our tests) can power the low-power version for $\sim 40$ days in continuous operation (assuming V$_\mathrm{in} = 2.5\,\mathrm{V}$) and the low-noise version for $\sim 3.5$ days.

\section{Non-local and high-impedance measurements} \label{sec:Results}
In order to demonstrate the use and capabilities of our apparatus we present two state-of-the-art measurements which would be adversely affected by artefacts in a normal experimental setup: the non-local resistance in a graphene Hall bar and the characterisation of a WS$_2$-based field-effect transistor with large contact resistance.

Non-local resistance measurements refer to the arrangement of the voltage probes far away from the current flow, typically in a Hall bar configuration. In recent years, this type of measurement in graphene devices has demonstrated several peculiarities of charge transport, such as: lifting of spin/valley degeneracy in monolayer graphene\cite{Abanin2011}, topological valley transport in bilayer graphene\cite{Shimazaki2015,Sui2015} and hydrodynamic transport of charges\cite{Levitov2016}. In all cases, it is necessary to resolve small currents and/or small voltages in environmentally sensitive devices and the preferred experimental technique relies on AC lock-in measurements. Here, we discuss on the artefacts that can be present in such measurements and how they can be removed with the use of the developed optocouplers. The device under examination is made of a single-layer exfoliated graphene flake encapsulated between two thin layers of hexagonal Boron Nitride (hBN) deposited on a Si substrate with $300\,\mathrm{nm}$ SiO$_2$ as gate dielectric. The graphene is contacted using Cr/Au electrodes in a side-contact configuration (or 1D contacts)\cite{Wang2013}. This structure and contact geometry allows the realisation of the ultra-high mobilities necessary to observe non-local effects in micron-scaled graphene devices\cite{Abanin2011,Levitov2016}.

\begin{figure*}{}
	\includegraphics[scale=1]{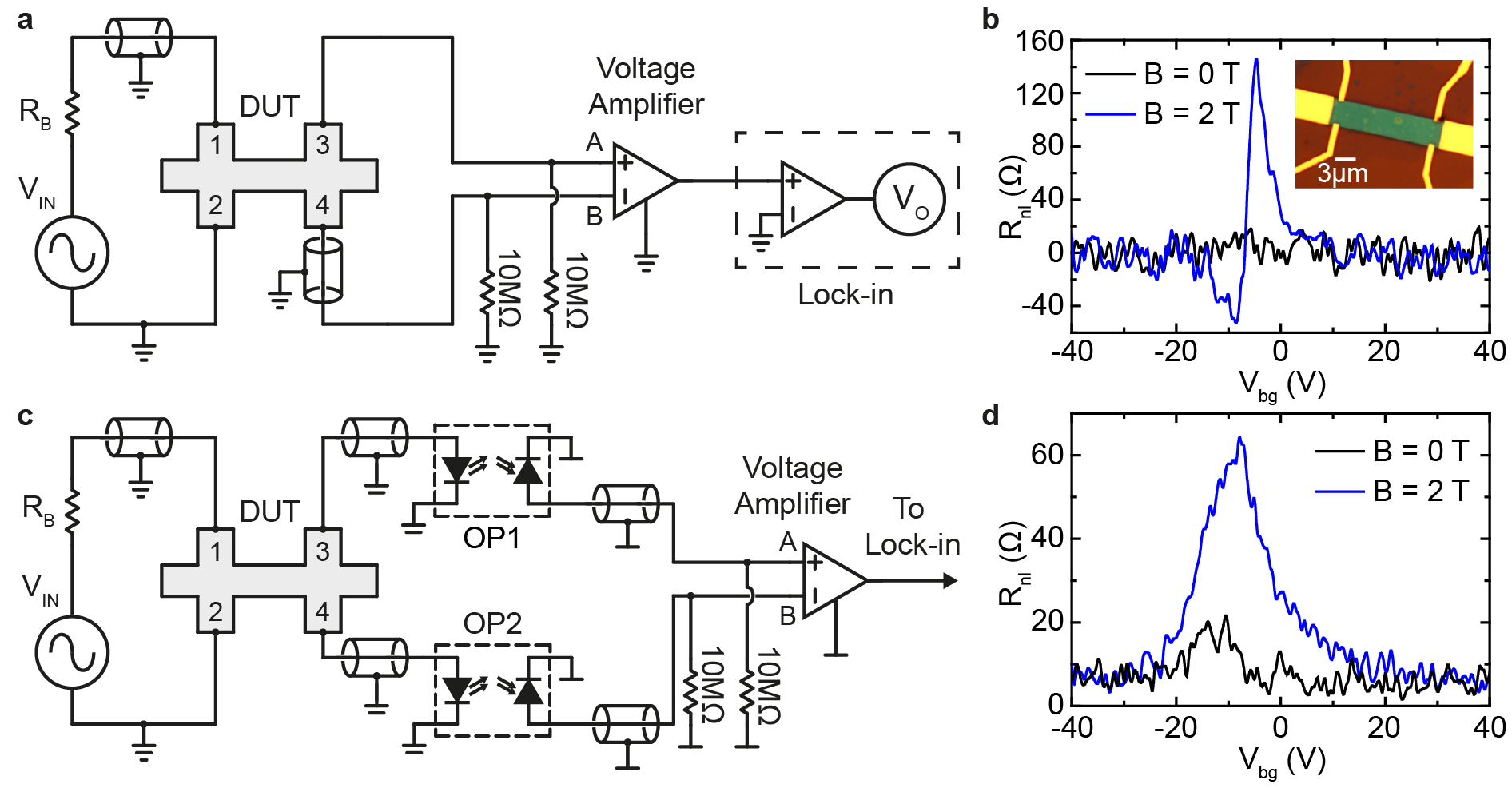}
	\caption{\label{Figure4}\textbf{Measurement of non-local resistance in graphene.} \textbf{a}, Standard lock-in measurement configuration. Current is injected between points 1 and 2 in a graphene Hall bar and the non-local voltage is measured between points 3 and 4. \textbf{b}, Non-local resistance ($\mathrm{R_{nl}}$) as a function of gate voltage in a sample device (inset) measured using the arrangement in panel \textbf{a} with and without magnetic field ($\mathrm{B=2\,T}$). \textbf{c}, Floating-probe arrangement with optocouplers of the same measurement and \textbf{d}, non-local resistance measured in this configuration.}
\end{figure*}

A typical room temperature non-local lock-in measurement of a graphene Hall bar is shown schematically in \cref{Figure4}a. The current is injected between contacts $1$ and $2$, through a known ballast resistor R$_B$ and the voltage drop (V$_\mathrm{nl}$) is measured between contacts $3$ and $4$. Such arrangement gives the results shown in \cref{Figure4}b, where the non-local resistance (R$_\mathrm{nl}$) is plotted as a function of gate voltage (V$_\mathrm{bg}$) in the presence, and absence, of a magnetic field (B) applied perpendicularly to the sample. We notice that the curve in \cref{Figure4}b for $\mathrm{B=2\,T}$ shows a negative resistance peak at V$_\mathrm{bg} = -10\,\mathrm{V}$, which corresponds to the charge neutrality point of the graphene Hall bar. Although the appearance of a negative bending resistance has been observed in graphene, and it is a signature of room temperature ballistic transport\cite{Mayorov2011,Banszerus2016}, such effect cannot be measured in our geometry because it requires a cross-shaped device, in which the current-injection (and voltage-sensing) is performed using pairs of orthogonal contacts. Therefore, the observed feature is an artefact introduced by the measurement configuration. This is caused by the incomplete rejection of the CMV (V$_\mathrm{cmv}$) at the input of the amplifier. A mismatch in the resistance between the current-injection leads produces a voltage drop which is then coupled to the voltage-probes via ground coupling and through the device itself (between pins $1-3$ and $2-4$). The spurious non-local voltage can be calculated as\cite{Sui2015}:

\begin{equation}
\label{eq:Voltage_CMV}
\mathrm{V_{nl}^{'}=\frac{V_{cmv}R_{in}}{R_{in}+R_1}-\frac{V_{cmv}R_{in}}{R_{in}+R_2}\approx(R_2-R_1)\frac{V_{cmv}}{R_{in}}},
\end{equation}
where R$_\mathrm{in}$ is the input impedance of the voltage amplifier, typically $1-10\,\mathrm{M\Omega}$, and R$_1$ (R$_2$) is the total resistance across the local part of the circuit, which include R$_\mathrm{B}$, the contact resistance of points $1$ and $3$ (or $2$ and $4$) and R$_\mathrm{in}$. The ground coupling is due to the dielectric leakage of the BNC cables towards the cable shielding, which effectively allows a signal generated between contacts $1$ and $2$ to be sensed between contacts $3$ and $4$ by-passing the device. The value of $\mathrm{V_{cmv}}$ increases with the channel resistance and it is maximum at the charge-neutrality point. Notably, depending on the values of R$_1$ and R$_2$, the value $\mathrm{V_{nl}^{'}}$ can be larger than the actual non-local signal and it can also have opposite sign (see \cref{Figure4}b). In order to suppress this spurious signal we adopted the configuration shown in \cref{Figure4}c. In this case two of the optocoupler circuits shown in \cref{Figure1} are used to float the voltage probes. The battery-powered circuits and the design of the measuring chamber allow to decouple at the same time the ground lines and the shells of the coaxial cables. The non-local measurement obtained using this arrangement is shown in \cref{Figure4}d, where no artefacts are present and a non-local resistance peak can be observed in the presence of a magnetic field at room temperature, as demonstrated in literature for a single-layer graphene on a Si/SiO$_2$ substrate, where such signal has been attributed to the flavour Hall effect (FHE)\cite{Abanin2011}. A residual non-local signal can be seen at zero magnetic field, which has been attributed in literature to joule heating\cite{Renard2014} (which causes a voltage in the non-local probes due to the mismatch in contact resistance). Notably our experiments show that the spurious signal can be as large as the actual non-local signal, as it can be seen by comparing the values of R$_\mathrm{nl}$ reported in \cref{Figure4}b and \cref{Figure4}d and calculated using \cref{eq:Voltage_CMV}. Furthermore, the ground coupling observed in the coaxial wires and connectors, responsible for the observed artefacts, is always present in a standard AC lock-in measurement, and it can be measured also in other Hi-Z devices, such as metallic thin films or in the insulating state of 2D materials.


\begin{figure}{}
	\includegraphics[scale=1]{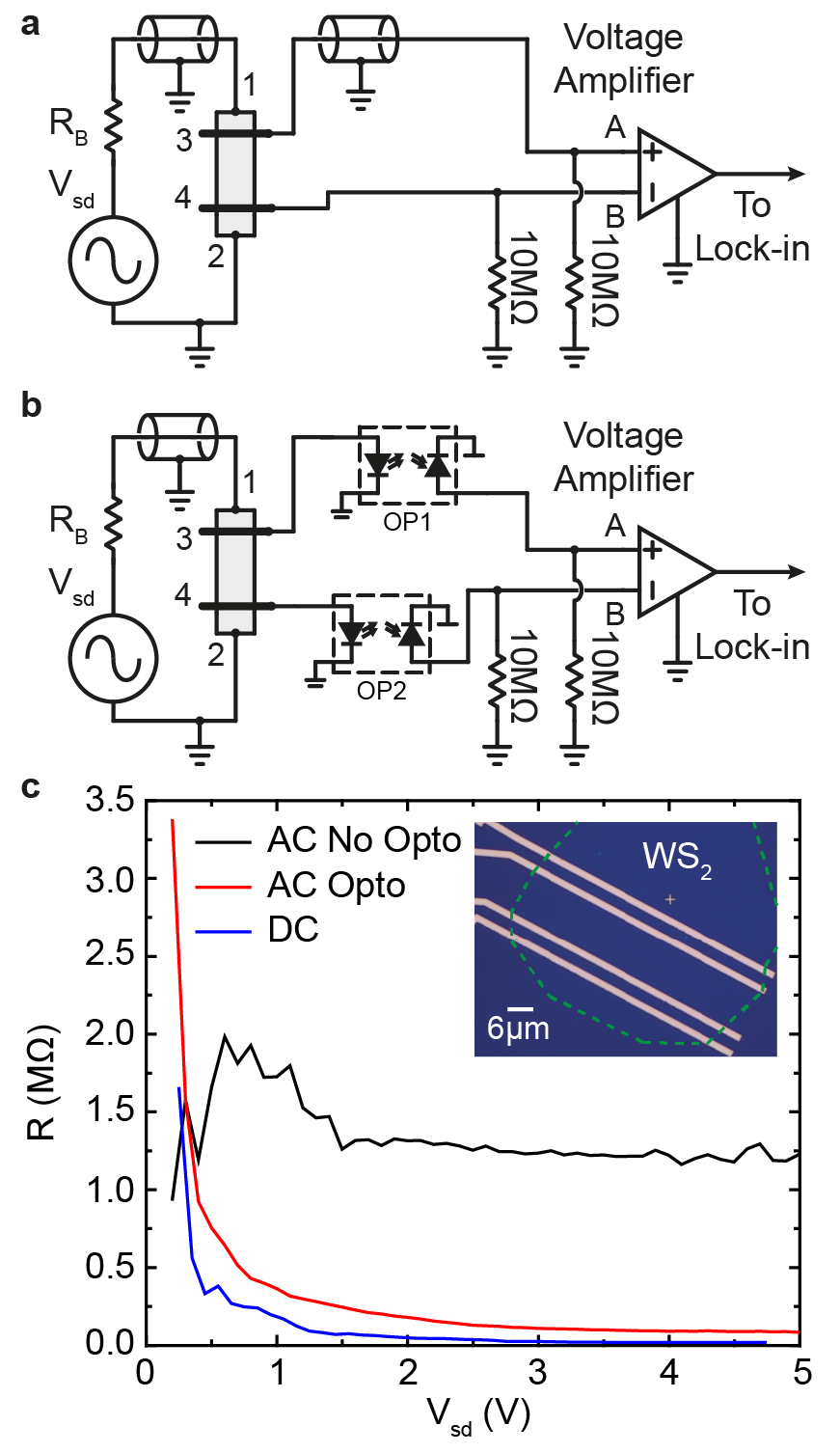}
	\caption{\label{Figure5}\textbf{Characterization of a WS$_2$ field-effect transistor.} \textbf{a}, Electrical circuit configuration for local 4-probe AC measurement of a field-effect transistor (FET). \textbf{b}, Floating-probe arrangement of the same FET measurement in panel \textbf{a}, coaxial cables are omitted for clarity. \textbf{c}, Resistance as a function of bias of a WS$_2$-based (FET), measured in the two AC configurations shown in panels \textbf{a} and \textbf{b} and in DC.}
\end{figure}

In the following we demonstrate the use of the developed optocoupler in the AC electrical characteristics of an atomically thin semiconductor transistor based on WS$_2$. The device is made of a single-layer chemical-vapour-deposition grown WS$_2$ on a Si substrate with $300\,\mathrm{nm}$ SiO$_2$ as gate dielectric, with lithographically defined Cr/Au electrodes. A micrograph of the device is shown in the inset of \cref{Figure5}c. The electrical circuit configuration used for the characterization without and with the optocouplers are shown in \cref{Figure5}a (b), respectively. The measurements obtained from the two different AC configurations are then benchmarked against a DC characterization of the device, see \cref{Figure5}c. Experimentally, we find that in the absence of the optocouplers, the AC characteristics differ significantly from the DC measurements. On the contrary, using the configuration shown in \cref{Figure5}b, the correct characteristic (\cref{Figure5}c, red line) is observed. In this case a sharp decrease of the channel resistance with applied bias voltage is expected in such material due to the large barrier formed at the contacts, a well-known issue in both exfoliated and CVD-grown transition-metal dichalcogenides\cite{Reale2017,DiBartolomeo2017}. Such characteristic is correctly displayed in the DC  measurement as well as in the AC configuration using the optical isolators.

\section{Conclusion} \label{sec:Conclusions}
In conclusion, we have presented an integrated opto-electrical circuit which allows the use of low-noise and low input impedance AC lock-in for the characterization of high-Z devices. We show that this optocoupler eliminates large spurious electrical signals by effectively decoupling the voltage sensing instruments from the current-driving circuitry. We demonstrate its performance in terms of fidelity, noise level, which is as low as the intrinsic noise level of the measuring instrument, and power consumption, which results in a continuous operation up to $40$ days using a small-size battery. The circuit can easily be miniaturized using surface-mount components and quickly up-scaled to fit multi-purposes applications when more than two contacts are needed in a small space, such as in modern experiments on quantum computing, large machines such as particle accelerators or in space applications. The developed optocouplers are highly suitable for the low-noise electrical characterization of narrow-gap states and semiconductors with potential applications in future quantum technologies such as computation, communication and sensing. Although in this work we focussed on the use of our circuitry to the characterization of 2D materials, the technique can be readily applied to the electrical measurement of a multitude of materials in which high-impedance states need to be accessed, such as polymers and composites, metalling thin-films, nanowires, nanotubes and quantum dots.

\section*{Supplementary Material}
Supplementary material is submitted with the current manuscript. Additional data, including EAGLE files for the schematics and PCB broads, related to this paper may be requested from the authors. Correspondence and requests for materials should be addressed to S.R.

\begin{acknowledgments}
J.D.M. acknowledges financial support from the Engineering and Physical Sciences Research Council (EPSRC) of the United Kingdom, via the EPSRC Centre for Doctoral Training in Metamaterials (Grant No. EP/L015331/1). S.F.R acknowledges financial support from the Higher Committee for Education Development in Iraq (HCED). S.R. and M.F.C. acknowledge financial support from EPSRC (Grant no. EP/K010050/1, EP/M001024/1, EP/M002438/1), from Royal Society international Exchanges Scheme 2016/R1, from The Leverhulme trust (grant title "Quantum Revolution").
\end{acknowledgments}

\section*{Author contributions}
A.D.S. conceived the idea, designed and characterized the circuit with the assistance of S.F.R. and C.P.T. A.W. and A.W. assisted in the design of the vacuum chamber and built the mechanical parts. J.D.M. fabricated the graphene devices. A.D.S. wrote the initial manuscript with input from all authors. M.F.C. and S.R. supervised the project and reviewed the manuscript.

\section*{Competing financial interests}
The authors declare no competing financial interests.

\clearpage

\bibliography{references_paperOptocouplers_AdeS}

\begin{thebibliography}{27}%
\makeatletter
\providecommand \@ifxundefined [1]{%
 \@ifx{#1\undefined}
}%
\providecommand \@ifnum [1]{%
 \ifnum #1\expandafter \@firstoftwo
 \else \expandafter \@secondoftwo
 \fi
}%
\providecommand \@ifx [1]{%
 \ifx #1\expandafter \@firstoftwo
 \else \expandafter \@secondoftwo
 \fi
}%
\providecommand \natexlab [1]{#1}%
\providecommand \enquote  [1]{``#1''}%
\providecommand \bibnamefont  [1]{#1}%
\providecommand \bibfnamefont [1]{#1}%
\providecommand \citenamefont [1]{#1}%
\providecommand \href@noop [0]{\@secondoftwo}%
\providecommand \href [0]{\begingroup \@sanitize@url \@href}%
\providecommand \@href[1]{\@@startlink{#1}\@@href}%
\providecommand \@@href[1]{\endgroup#1\@@endlink}%
\providecommand \@sanitize@url [0]{\catcode `\\12\catcode `\$12\catcode
  `\&12\catcode `\#12\catcode `\^12\catcode `\_12\catcode `\%12\relax}%
\providecommand \@@startlink[1]{}%
\providecommand \@@endlink[0]{}%
\providecommand \url  [0]{\begingroup\@sanitize@url \@url }%
\providecommand \@url [1]{\endgroup\@href {#1}{\urlprefix }}%
\providecommand \urlprefix  [0]{URL }%
\providecommand \Eprint [0]{\href }%
\providecommand \doibase [0]{http://dx.doi.org/}%
\providecommand \selectlanguage [0]{\@gobble}%
\providecommand \bibinfo  [0]{\@secondoftwo}%
\providecommand \bibfield  [0]{\@secondoftwo}%
\providecommand \translation [1]{[#1]}%
\providecommand \BibitemOpen [0]{}%
\providecommand \bibitemStop [0]{}%
\providecommand \bibitemNoStop [0]{.\EOS\space}%
\providecommand \EOS [0]{\spacefactor3000\relax}%
\providecommand \BibitemShut  [1]{\csname bibitem#1\endcsname}%
\let\auto@bib@innerbib\@empty
\bibitem [{\citenamefont {{Kamerlingh Onnes}}(1911)}]{KamerlinghOnnes1911}%
  \BibitemOpen
  \bibfield  {author} {\bibinfo {author} {\bibfnamefont {H.}~\bibnamefont
  {{Kamerlingh Onnes}}},\ }in\ \href
  {http://www.dwc.knaw.nl/DL/publications/PU00013124.pdf} {\emph {\bibinfo
  {booktitle} {KNAW, Proceedings}}},\ Vol.\ \bibinfo {volume} {14 I}\ (\bibinfo
  {year} {1911})\ pp.\ \bibinfo {pages} {113--115}\BibitemShut {NoStop}%
\bibitem [{\citenamefont {Geim}\ and\ \citenamefont
  {Novoselov}(2007)}]{Geim2007}%
  \BibitemOpen
  \bibfield  {author} {\bibinfo {author} {\bibfnamefont {A.~K.}\ \bibnamefont
  {Geim}}\ and\ \bibinfo {author} {\bibfnamefont {K.~S.}\ \bibnamefont
  {Novoselov}},\ }\href {\doibase 10.1038/nmat1849} {\bibfield  {journal}
  {\bibinfo  {journal} {Nat Mater}\ }\textbf {\bibinfo {volume} {6}},\ \bibinfo
  {pages} {183} (\bibinfo {year} {2007})}\BibitemShut {NoStop}%
\bibitem [{\citenamefont {Wang}\ \emph {et~al.}(2012)\citenamefont {Wang},
  \citenamefont {Kalantar-Zadeh}, \citenamefont {Kis}, \citenamefont
  {Coleman},\ and\ \citenamefont {Strano}}]{Wang2012}%
  \BibitemOpen
  \bibfield  {author} {\bibinfo {author} {\bibfnamefont {Q.~H.}\ \bibnamefont
  {Wang}}, \bibinfo {author} {\bibfnamefont {K.}~\bibnamefont
  {Kalantar-Zadeh}}, \bibinfo {author} {\bibfnamefont {A.}~\bibnamefont {Kis}},
  \bibinfo {author} {\bibfnamefont {J.~N.}\ \bibnamefont {Coleman}}, \ and\
  \bibinfo {author} {\bibfnamefont {M.~S.}\ \bibnamefont {Strano}},\ }\href
  {\doibase 10.1038/nnano.2012.193} {\bibfield  {journal} {\bibinfo  {journal}
  {Nat Nano}\ }\textbf {\bibinfo {volume} {7}},\ \bibinfo {pages} {699}
  (\bibinfo {year} {2012})}\BibitemShut {NoStop}%
\bibitem [{\citenamefont {Chhowalla}\ \emph {et~al.}(2013)\citenamefont
  {Chhowalla}, \citenamefont {Shin}, \citenamefont {Eda}, \citenamefont {Li},
  \citenamefont {Loh},\ and\ \citenamefont {Zhang}}]{Chhowalla2013}%
  \BibitemOpen
  \bibfield  {author} {\bibinfo {author} {\bibfnamefont {M.}~\bibnamefont
  {Chhowalla}}, \bibinfo {author} {\bibfnamefont {H.~S.}\ \bibnamefont {Shin}},
  \bibinfo {author} {\bibfnamefont {G.}~\bibnamefont {Eda}}, \bibinfo {author}
  {\bibfnamefont {L.-J.}\ \bibnamefont {Li}}, \bibinfo {author} {\bibfnamefont
  {K.~P.}\ \bibnamefont {Loh}}, \ and\ \bibinfo {author} {\bibfnamefont
  {H.}~\bibnamefont {Zhang}},\ }\href {\doibase 10.1038/nchem.1589} {\bibfield
  {journal} {\bibinfo  {journal} {Nat Chem}\ }\textbf {\bibinfo {volume} {5}},\
  \bibinfo {pages} {263} (\bibinfo {year} {2013})}\BibitemShut {NoStop}%
\bibitem [{\citenamefont {Kosterlitz}(2016)}]{Kosterlitz2016}%
  \BibitemOpen
  \bibfield  {author} {\bibinfo {author} {\bibfnamefont {J.~M.}\ \bibnamefont
  {Kosterlitz}},\ }\href {http://stacks.iop.org/0034-4885/79/i=2/a=026001}
  {\bibfield  {journal} {\bibinfo  {journal} {Reports on Progress in Physics}\
  }\textbf {\bibinfo {volume} {79}},\ \bibinfo {pages} {026001} (\bibinfo
  {year} {2016})}\BibitemShut {NoStop}%
\bibitem [{\citenamefont {Qi}\ and\ \citenamefont {Zhang}(2011)}]{Qi2011}%
  \BibitemOpen
  \bibfield  {author} {\bibinfo {author} {\bibfnamefont {X.-L.}\ \bibnamefont
  {Qi}}\ and\ \bibinfo {author} {\bibfnamefont {S.-C.}\ \bibnamefont {Zhang}},\
  }\href {\doibase 10.1103/RevModPhys.83.1057} {\bibfield  {journal} {\bibinfo
  {journal} {Rev. Mod. Phys.}\ }\textbf {\bibinfo {volume} {83}},\ \bibinfo
  {pages} {1057} (\bibinfo {year} {2011})}\BibitemShut {NoStop}%
\bibitem [{\citenamefont {Yuasa}\ \emph {et~al.}(2004)\citenamefont {Yuasa},
  \citenamefont {Nagahama}, \citenamefont {Fukushima}, \citenamefont {Suzuki},\
  and\ \citenamefont {Ando}}]{Yuasa2004}%
  \BibitemOpen
  \bibfield  {author} {\bibinfo {author} {\bibfnamefont {S.}~\bibnamefont
  {Yuasa}}, \bibinfo {author} {\bibfnamefont {T.}~\bibnamefont {Nagahama}},
  \bibinfo {author} {\bibfnamefont {A.}~\bibnamefont {Fukushima}}, \bibinfo
  {author} {\bibfnamefont {Y.}~\bibnamefont {Suzuki}}, \ and\ \bibinfo {author}
  {\bibfnamefont {K.}~\bibnamefont {Ando}},\ }\href {\doibase 10.1038/nmat1257}
  {\bibfield  {journal} {\bibinfo  {journal} {Nat Mater}\ }\textbf {\bibinfo
  {volume} {3}},\ \bibinfo {pages} {868} (\bibinfo {year} {2004})}\BibitemShut
  {NoStop}%
\bibitem [{\citenamefont {Abanin}\ \emph {et~al.}(2011)\citenamefont {Abanin},
  \citenamefont {Morozov}, \citenamefont {Ponomarenko}, \citenamefont
  {Gorbachev}, \citenamefont {Mayorov}, \citenamefont {Katsnelson},
  \citenamefont {Watanabe}, \citenamefont {Taniguchi}, \citenamefont
  {Novoselov}, \citenamefont {Levitov},\ and\ \citenamefont
  {Geim}}]{Abanin2011}%
  \BibitemOpen
  \bibfield  {author} {\bibinfo {author} {\bibfnamefont {D.~A.}\ \bibnamefont
  {Abanin}}, \bibinfo {author} {\bibfnamefont {S.~V.}\ \bibnamefont {Morozov}},
  \bibinfo {author} {\bibfnamefont {L.~A.}\ \bibnamefont {Ponomarenko}},
  \bibinfo {author} {\bibfnamefont {R.~V.}\ \bibnamefont {Gorbachev}}, \bibinfo
  {author} {\bibfnamefont {A.~S.}\ \bibnamefont {Mayorov}}, \bibinfo {author}
  {\bibfnamefont {M.~I.}\ \bibnamefont {Katsnelson}}, \bibinfo {author}
  {\bibfnamefont {K.}~\bibnamefont {Watanabe}}, \bibinfo {author}
  {\bibfnamefont {T.}~\bibnamefont {Taniguchi}}, \bibinfo {author}
  {\bibfnamefont {K.~S.}\ \bibnamefont {Novoselov}}, \bibinfo {author}
  {\bibfnamefont {L.~S.}\ \bibnamefont {Levitov}}, \ and\ \bibinfo {author}
  {\bibfnamefont {A.~K.}\ \bibnamefont {Geim}},\ }\href {\doibase
  10.1126/science.1199595} {\bibfield  {journal} {\bibinfo  {journal}
  {Science}\ }\textbf {\bibinfo {volume} {332}},\ \bibinfo {pages} {328}
  (\bibinfo {year} {2011})}\BibitemShut {NoStop}%
\bibitem [{\citenamefont {Bandurin}\ \emph {et~al.}(2016)\citenamefont
  {Bandurin}, \citenamefont {Torre}, \citenamefont {Kumar}, \citenamefont
  {Ben~Shalom}, \citenamefont {Tomadin}, \citenamefont {Principi},
  \citenamefont {Auton}, \citenamefont {Khestanova}, \citenamefont {Novoselov},
  \citenamefont {Grigorieva}, \citenamefont {Ponomarenko}, \citenamefont
  {Geim},\ and\ \citenamefont {Polini}}]{Bandurinaad2016}%
  \BibitemOpen
  \bibfield  {author} {\bibinfo {author} {\bibfnamefont {D.~A.}\ \bibnamefont
  {Bandurin}}, \bibinfo {author} {\bibfnamefont {I.}~\bibnamefont {Torre}},
  \bibinfo {author} {\bibfnamefont {R.~K.}\ \bibnamefont {Kumar}}, \bibinfo
  {author} {\bibfnamefont {M.}~\bibnamefont {Ben~Shalom}}, \bibinfo {author}
  {\bibfnamefont {A.}~\bibnamefont {Tomadin}}, \bibinfo {author} {\bibfnamefont
  {A.}~\bibnamefont {Principi}}, \bibinfo {author} {\bibfnamefont {G.~H.}\
  \bibnamefont {Auton}}, \bibinfo {author} {\bibfnamefont {E.}~\bibnamefont
  {Khestanova}}, \bibinfo {author} {\bibfnamefont {K.~S.}\ \bibnamefont
  {Novoselov}}, \bibinfo {author} {\bibfnamefont {I.~V.}\ \bibnamefont
  {Grigorieva}}, \bibinfo {author} {\bibfnamefont {L.~A.}\ \bibnamefont
  {Ponomarenko}}, \bibinfo {author} {\bibfnamefont {A.~K.}\ \bibnamefont
  {Geim}}, \ and\ \bibinfo {author} {\bibfnamefont {M.}~\bibnamefont
  {Polini}},\ }\href {\doibase 10.1126/science.aad0201} {\bibfield  {journal}
  {\bibinfo  {journal} {Science}\ }\textbf {\bibinfo {volume} {351}},\ \bibinfo
  {pages} {1055} (\bibinfo {year} {2016})}\BibitemShut {NoStop}%
\bibitem [{\citenamefont {Craciun}\ \emph {et~al.}(2009)\citenamefont
  {Craciun}, \citenamefont {Russo}, \citenamefont {Yamamoto}, \citenamefont
  {Oostinga}, \citenamefont {Morpurgo},\ and\ \citenamefont
  {Tarucha}}]{Craciun2009}%
  \BibitemOpen
  \bibfield  {author} {\bibinfo {author} {\bibfnamefont {M.~F.}\ \bibnamefont
  {Craciun}}, \bibinfo {author} {\bibfnamefont {S.}~\bibnamefont {Russo}},
  \bibinfo {author} {\bibfnamefont {M.}~\bibnamefont {Yamamoto}}, \bibinfo
  {author} {\bibfnamefont {J.~B.}\ \bibnamefont {Oostinga}}, \bibinfo {author}
  {\bibfnamefont {A.~F.}\ \bibnamefont {Morpurgo}}, \ and\ \bibinfo {author}
  {\bibfnamefont {S.}~\bibnamefont {Tarucha}},\ }\href {\doibase
  10.1038/nnano.2009.89} {\bibfield  {journal} {\bibinfo  {journal} {Nature
  Nanotechnology}\ }\textbf {\bibinfo {volume} {4}},\ \bibinfo {pages} {383}
  (\bibinfo {year} {2009})}\BibitemShut {NoStop}%
\bibitem [{\citenamefont {Goovaerts}\ \emph {et~al.}(1998)\citenamefont
  {Goovaerts}, \citenamefont {Faes}, \citenamefont {Raaijmakers},\ and\
  \citenamefont {Heethaar}}]{Goovaerts1998}%
  \BibitemOpen
  \bibfield  {author} {\bibinfo {author} {\bibfnamefont {H.~G.}\ \bibnamefont
  {Goovaerts}}, \bibinfo {author} {\bibfnamefont {T.~J.~C.}\ \bibnamefont
  {Faes}}, \bibinfo {author} {\bibfnamefont {E.}~\bibnamefont {Raaijmakers}}, \
  and\ \bibinfo {author} {\bibfnamefont {R.~M.}\ \bibnamefont {Heethaar}},\
  }\href {\doibase 10.1007/BF02518881} {\bibfield  {journal} {\bibinfo
  {journal} {Medical and Biological Engineering and Computing}\ }\textbf
  {\bibinfo {volume} {36}},\ \bibinfo {pages} {761} (\bibinfo {year}
  {1998})}\BibitemShut {NoStop}%
\bibitem [{\citenamefont {Kotler}\ \emph {et~al.}(2011)\citenamefont {Kotler},
  \citenamefont {Akerman}, \citenamefont {Glickman}, \citenamefont {Keselman},\
  and\ \citenamefont {Ozeri}}]{Kotler2011}%
  \BibitemOpen
  \bibfield  {author} {\bibinfo {author} {\bibfnamefont {S.}~\bibnamefont
  {Kotler}}, \bibinfo {author} {\bibfnamefont {N.}~\bibnamefont {Akerman}},
  \bibinfo {author} {\bibfnamefont {Y.}~\bibnamefont {Glickman}}, \bibinfo
  {author} {\bibfnamefont {A.}~\bibnamefont {Keselman}}, \ and\ \bibinfo
  {author} {\bibfnamefont {R.}~\bibnamefont {Ozeri}},\ }\href {\doibase
  10.1038/nature10010} {\bibfield  {journal} {\bibinfo  {journal} {Nature}\
  }\textbf {\bibinfo {volume} {473}},\ \bibinfo {pages} {61} (\bibinfo {year}
  {2011})}\BibitemShut {NoStop}%
\bibitem [{\citenamefont {Sui}\ \emph {et~al.}(2015)\citenamefont {Sui},
  \citenamefont {Chen}, \citenamefont {Ma}, \citenamefont {Shan}, \citenamefont
  {Tian}, \citenamefont {Watanabe}, \citenamefont {Taniguchi}, \citenamefont
  {Jin}, \citenamefont {Yao}, \citenamefont {Xiao},\ and\ \citenamefont
  {Zhang}}]{Sui2015}%
  \BibitemOpen
  \bibfield  {author} {\bibinfo {author} {\bibfnamefont {M.}~\bibnamefont
  {Sui}}, \bibinfo {author} {\bibfnamefont {G.}~\bibnamefont {Chen}}, \bibinfo
  {author} {\bibfnamefont {L.}~\bibnamefont {Ma}}, \bibinfo {author}
  {\bibfnamefont {W.-Y.}\ \bibnamefont {Shan}}, \bibinfo {author}
  {\bibfnamefont {D.}~\bibnamefont {Tian}}, \bibinfo {author} {\bibfnamefont
  {K.}~\bibnamefont {Watanabe}}, \bibinfo {author} {\bibfnamefont
  {T.}~\bibnamefont {Taniguchi}}, \bibinfo {author} {\bibfnamefont
  {X.}~\bibnamefont {Jin}}, \bibinfo {author} {\bibfnamefont {W.}~\bibnamefont
  {Yao}}, \bibinfo {author} {\bibfnamefont {D.}~\bibnamefont {Xiao}}, \ and\
  \bibinfo {author} {\bibfnamefont {Y.}~\bibnamefont {Zhang}},\ }\href
  {http://dx.doi.org/10.1038/nphys3485} {\bibfield  {journal} {\bibinfo
  {journal} {Nat Phys}\ }\textbf {\bibinfo {volume} {11}},\ \bibinfo {pages}
  {1027} (\bibinfo {year} {2015})}\BibitemShut {NoStop}%
\bibitem [{\citenamefont {Shimazaki}\ \emph {et~al.}(2015)\citenamefont
  {Shimazaki}, \citenamefont {Yamamoto}, \citenamefont {Borzenets},
  \citenamefont {Watanabe}, \citenamefont {Taniguchi},\ and\ \citenamefont
  {Tarucha}}]{Shimazaki2015}%
  \BibitemOpen
  \bibfield  {author} {\bibinfo {author} {\bibfnamefont {Y.}~\bibnamefont
  {Shimazaki}}, \bibinfo {author} {\bibfnamefont {M.}~\bibnamefont {Yamamoto}},
  \bibinfo {author} {\bibfnamefont {I.~V.}\ \bibnamefont {Borzenets}}, \bibinfo
  {author} {\bibfnamefont {K.}~\bibnamefont {Watanabe}}, \bibinfo {author}
  {\bibfnamefont {T.}~\bibnamefont {Taniguchi}}, \ and\ \bibinfo {author}
  {\bibfnamefont {S.}~\bibnamefont {Tarucha}},\ }\href
  {http://dx.doi.org/10.1038/nphys3551} {\bibfield  {journal} {\bibinfo
  {journal} {Nat Phys}\ }\textbf {\bibinfo {volume} {11}},\ \bibinfo {pages}
  {1032} (\bibinfo {year} {2015})}\BibitemShut {NoStop}%
\bibitem [{\citenamefont {Overweg}\ \emph {et~al.}(2017)\citenamefont
  {Overweg}, \citenamefont {Eggimann}, \citenamefont {Eich}, \citenamefont
  {Pisoni}, \citenamefont {Lee}, \citenamefont {Rickhaus}, \citenamefont {Ihn},
  \citenamefont {Ensslin}, \citenamefont {Chen}, \citenamefont {Slizovskiy},
  \citenamefont {{Fal 'ko}}, \citenamefont {Watanabe},\ and\ \citenamefont
  {Taniguchi}}]{Overweg2017}%
  \BibitemOpen
  \bibfield  {author} {\bibinfo {author} {\bibfnamefont {H.}~\bibnamefont
  {Overweg}}, \bibinfo {author} {\bibfnamefont {H.}~\bibnamefont {Eggimann}},
  \bibinfo {author} {\bibfnamefont {M.}~\bibnamefont {Eich}}, \bibinfo {author}
  {\bibfnamefont {R.}~\bibnamefont {Pisoni}}, \bibinfo {author} {\bibfnamefont
  {Y.}~\bibnamefont {Lee}}, \bibinfo {author} {\bibfnamefont {P.}~\bibnamefont
  {Rickhaus}}, \bibinfo {author} {\bibfnamefont {T.}~\bibnamefont {Ihn}},
  \bibinfo {author} {\bibfnamefont {K.}~\bibnamefont {Ensslin}}, \bibinfo
  {author} {\bibfnamefont {X.}~\bibnamefont {Chen}}, \bibinfo {author}
  {\bibfnamefont {S.}~\bibnamefont {Slizovskiy}}, \bibinfo {author}
  {\bibfnamefont {V.}~\bibnamefont {{Fal 'ko}}}, \bibinfo {author}
  {\bibfnamefont {K.}~\bibnamefont {Watanabe}}, \ and\ \bibinfo {author}
  {\bibfnamefont {T.}~\bibnamefont {Taniguchi}},\ }\href@noop {} {\bibfield
  {journal} {\bibinfo  {journal} {arxiv.org/pdf/1707.09282.pdf}\ } (\bibinfo
  {year} {2017})}\BibitemShut {NoStop}%
\bibitem [{\citenamefont {Morrison}(1998)}]{Morrison_GND}%
  \BibitemOpen
  \bibfield  {author} {\bibinfo {author} {\bibfnamefont {R.}~\bibnamefont
  {Morrison}},\ }\href@noop {} {\emph {\bibinfo {title} {Grounding and
  shielding techniques}}},\ \bibinfo {edition} {4th}\ ed.,\ edited by\ \bibinfo
  {editor} {\bibnamefont {Wiley}}\ (\bibinfo  {publisher} {John Wiley and Sons,
  Inc.},\ \bibinfo {year} {1998})\ Chap.~\bibinfo {chapter} {4}, pp.\ \bibinfo
  {pages} {73--74}\BibitemShut {NoStop}%
\bibitem [{\citenamefont {{Agilent (Hewlett-Packard)}}(2005)}]{HCNR_Datasheet}%
  \BibitemOpen
  \bibfield  {author} {\bibinfo {author} {\bibnamefont {{Agilent
  (Hewlett-Packard)}}},\ }\href
  {https://upverter.com/datasheet/f9b1f81ec2e8a843f62c44e1b1d173bf8e17cd3f.pdf}
  {\emph {\bibinfo {title} {HCNR200/1 High-Linearity Analog Optocouplers
  Datasheet}}} (\bibinfo {year} {2005})\BibitemShut {NoStop}%
\bibitem [{\citenamefont {Khodkov}\ \emph {et~al.}(2015)\citenamefont
  {Khodkov}, \citenamefont {Khrapach}, \citenamefont {Craciun},\ and\
  \citenamefont {Russo}}]{Khodkov2015}%
  \BibitemOpen
  \bibfield  {author} {\bibinfo {author} {\bibfnamefont {T.}~\bibnamefont
  {Khodkov}}, \bibinfo {author} {\bibfnamefont {I.}~\bibnamefont {Khrapach}},
  \bibinfo {author} {\bibfnamefont {M.~F.}\ \bibnamefont {Craciun}}, \ and\
  \bibinfo {author} {\bibfnamefont {S.}~\bibnamefont {Russo}},\ }\href
  {\doibase 10.1021/acs.nanolett.5b00772} {\bibfield  {journal} {\bibinfo
  {journal} {Nano Letters}\ }\textbf {\bibinfo {volume} {15}},\ \bibinfo
  {pages} {4429} (\bibinfo {year} {2015})}\BibitemShut {NoStop}%
\bibitem [{\citenamefont {{De Sanctis}}\ \emph {et~al.}(2017)\citenamefont {{De
  Sanctis}}, \citenamefont {Jones}, \citenamefont {Townsend}, \citenamefont
  {Craciun},\ and\ \citenamefont {Russo}}]{DeSanctis2017_G16}%
  \BibitemOpen
  \bibfield  {author} {\bibinfo {author} {\bibfnamefont {A.}~\bibnamefont {{De
  Sanctis}}}, \bibinfo {author} {\bibfnamefont {G.~F.}\ \bibnamefont {Jones}},
  \bibinfo {author} {\bibfnamefont {N.~J.}\ \bibnamefont {Townsend}}, \bibinfo
  {author} {\bibfnamefont {M.~F.}\ \bibnamefont {Craciun}}, \ and\ \bibinfo
  {author} {\bibfnamefont {S.}~\bibnamefont {Russo}},\ }\href {\doibase
  10.1063/1.4982358} {\bibfield  {journal} {\bibinfo  {journal} {Review of
  Scientific Instruments}\ }\textbf {\bibinfo {volume} {88}},\ \bibinfo {pages}
  {055102} (\bibinfo {year} {2017})}\BibitemShut {NoStop}%
\bibitem [{\citenamefont {Asparuhova}\ and\ \citenamefont
  {Gadjeva}(2004)}]{Asparuhova2004}%
  \BibitemOpen
  \bibfield  {author} {\bibinfo {author} {\bibfnamefont {K.~K.}\ \bibnamefont
  {Asparuhova}}\ and\ \bibinfo {author} {\bibfnamefont {E.~D.}\ \bibnamefont
  {Gadjeva}},\ }in\ \href {\doibase 10.1109/ISSE.2004.1490859} {\emph {\bibinfo
  {booktitle} {27th International Spring Seminar on Electronics Technology:
  Meeting the Challenges of Electronics Technology Progress}}},\ Vol.~\bibinfo
  {volume} {3}\ (\bibinfo {year} {2004})\ pp.\ \bibinfo {pages}
  {471--475}\BibitemShut {NoStop}%
\bibitem [{\citenamefont {Levitov}\ and\ \citenamefont
  {Falkovich}(2016)}]{Levitov2016}%
  \BibitemOpen
  \bibfield  {author} {\bibinfo {author} {\bibfnamefont {L.}~\bibnamefont
  {Levitov}}\ and\ \bibinfo {author} {\bibfnamefont {G.}~\bibnamefont
  {Falkovich}},\ }\href {http://dx.doi.org/10.1038/nphys3667} {\bibfield
  {journal} {\bibinfo  {journal} {Nat Phys}\ }\textbf {\bibinfo {volume}
  {12}},\ \bibinfo {pages} {672} (\bibinfo {year} {2016})}\BibitemShut
  {NoStop}%
\bibitem [{\citenamefont {Wang}\ \emph {et~al.}(2013)\citenamefont {Wang},
  \citenamefont {Meric}, \citenamefont {Huang}, \citenamefont {Gao},
  \citenamefont {Gao}, \citenamefont {Tran}, \citenamefont {Taniguchi},
  \citenamefont {Watanabe}, \citenamefont {Campos}, \citenamefont {Muller},
  \citenamefont {Guo}, \citenamefont {Kim}, \citenamefont {Hone}, \citenamefont
  {Shepard},\ and\ \citenamefont {Dean}}]{Wang2013}%
  \BibitemOpen
  \bibfield  {author} {\bibinfo {author} {\bibfnamefont {L.}~\bibnamefont
  {Wang}}, \bibinfo {author} {\bibfnamefont {I.}~\bibnamefont {Meric}},
  \bibinfo {author} {\bibfnamefont {P.~Y.}\ \bibnamefont {Huang}}, \bibinfo
  {author} {\bibfnamefont {Q.}~\bibnamefont {Gao}}, \bibinfo {author}
  {\bibfnamefont {Y.}~\bibnamefont {Gao}}, \bibinfo {author} {\bibfnamefont
  {H.}~\bibnamefont {Tran}}, \bibinfo {author} {\bibfnamefont {T.}~\bibnamefont
  {Taniguchi}}, \bibinfo {author} {\bibfnamefont {K.}~\bibnamefont {Watanabe}},
  \bibinfo {author} {\bibfnamefont {L.~M.}\ \bibnamefont {Campos}}, \bibinfo
  {author} {\bibfnamefont {D.~A.}\ \bibnamefont {Muller}}, \bibinfo {author}
  {\bibfnamefont {J.}~\bibnamefont {Guo}}, \bibinfo {author} {\bibfnamefont
  {P.}~\bibnamefont {Kim}}, \bibinfo {author} {\bibfnamefont {J.}~\bibnamefont
  {Hone}}, \bibinfo {author} {\bibfnamefont {K.~L.}\ \bibnamefont {Shepard}}, \
  and\ \bibinfo {author} {\bibfnamefont {C.~R.}\ \bibnamefont {Dean}},\ }\href
  {\doibase 10.1126/science.1244358} {\bibfield  {journal} {\bibinfo  {journal}
  {Science}\ }\textbf {\bibinfo {volume} {342}},\ \bibinfo {pages} {614}
  (\bibinfo {year} {2013})}\BibitemShut {NoStop}%
\bibitem [{\citenamefont {Mayorov}\ \emph {et~al.}(2011)\citenamefont
  {Mayorov}, \citenamefont {Gorbachev}, \citenamefont {Morozov}, \citenamefont
  {Britnell}, \citenamefont {Jalil}, \citenamefont {Ponomarenko}, \citenamefont
  {Blake}, \citenamefont {Novoselov}, \citenamefont {Watanabe}, \citenamefont
  {Taniguchi},\ and\ \citenamefont {Geim}}]{Mayorov2011}%
  \BibitemOpen
  \bibfield  {author} {\bibinfo {author} {\bibfnamefont {A.~S.}\ \bibnamefont
  {Mayorov}}, \bibinfo {author} {\bibfnamefont {R.~V.}\ \bibnamefont
  {Gorbachev}}, \bibinfo {author} {\bibfnamefont {S.~V.}\ \bibnamefont
  {Morozov}}, \bibinfo {author} {\bibfnamefont {L.}~\bibnamefont {Britnell}},
  \bibinfo {author} {\bibfnamefont {R.}~\bibnamefont {Jalil}}, \bibinfo
  {author} {\bibfnamefont {L.~A.}\ \bibnamefont {Ponomarenko}}, \bibinfo
  {author} {\bibfnamefont {P.}~\bibnamefont {Blake}}, \bibinfo {author}
  {\bibfnamefont {K.~S.}\ \bibnamefont {Novoselov}}, \bibinfo {author}
  {\bibfnamefont {K.}~\bibnamefont {Watanabe}}, \bibinfo {author}
  {\bibfnamefont {T.}~\bibnamefont {Taniguchi}}, \ and\ \bibinfo {author}
  {\bibfnamefont {A.~K.}\ \bibnamefont {Geim}},\ }\href {\doibase
  10.1021/nl200758b} {\bibfield  {journal} {\bibinfo  {journal} {Nano Letters}\
  }\textbf {\bibinfo {volume} {11}},\ \bibinfo {pages} {2396} (\bibinfo {year}
  {2011})}\BibitemShut {NoStop}%
\bibitem [{\citenamefont {Banszerus}\ \emph {et~al.}(2016)\citenamefont
  {Banszerus}, \citenamefont {Schmitz}, \citenamefont {Engels}, \citenamefont
  {Goldsche}, \citenamefont {Watanabe}, \citenamefont {Taniguchi},
  \citenamefont {Beschoten},\ and\ \citenamefont {Stampfer}}]{Banszerus2016}%
  \BibitemOpen
  \bibfield  {author} {\bibinfo {author} {\bibfnamefont {L.}~\bibnamefont
  {Banszerus}}, \bibinfo {author} {\bibfnamefont {M.}~\bibnamefont {Schmitz}},
  \bibinfo {author} {\bibfnamefont {S.}~\bibnamefont {Engels}}, \bibinfo
  {author} {\bibfnamefont {M.}~\bibnamefont {Goldsche}}, \bibinfo {author}
  {\bibfnamefont {K.}~\bibnamefont {Watanabe}}, \bibinfo {author}
  {\bibfnamefont {T.}~\bibnamefont {Taniguchi}}, \bibinfo {author}
  {\bibfnamefont {B.}~\bibnamefont {Beschoten}}, \ and\ \bibinfo {author}
  {\bibfnamefont {C.}~\bibnamefont {Stampfer}},\ }\href {\doibase
  10.1021/acs.nanolett.5b04840} {\bibfield  {journal} {\bibinfo  {journal}
  {Nano Letters}\ }\textbf {\bibinfo {volume} {16}},\ \bibinfo {pages} {1387}
  (\bibinfo {year} {2016})}\BibitemShut {NoStop}%
\bibitem [{\citenamefont {Renard}, \citenamefont {Studer},\ and\ \citenamefont
  {Folk}(2014)}]{Renard2014}%
  \BibitemOpen
  \bibfield  {author} {\bibinfo {author} {\bibfnamefont {J.}~\bibnamefont
  {Renard}}, \bibinfo {author} {\bibfnamefont {M.}~\bibnamefont {Studer}}, \
  and\ \bibinfo {author} {\bibfnamefont {J.~A.}\ \bibnamefont {Folk}},\ }\href
  {\doibase 10.1103/PhysRevLett.112.116601} {\bibfield  {journal} {\bibinfo
  {journal} {Physical Review Letters}\ }\textbf {\bibinfo {volume} {112}},\
  \bibinfo {pages} {1} (\bibinfo {year} {2014})},\ \Eprint
  {http://arxiv.org/abs/1309.7016} {1309.7016} \BibitemShut {NoStop}%
\bibitem [{\citenamefont {Reale}\ \emph {et~al.}(2017)\citenamefont {Reale},
  \citenamefont {Palczynski}, \citenamefont {Amit}, \citenamefont {Jones},
  \citenamefont {Mehew}, \citenamefont {Bacon}, \citenamefont {Ni},
  \citenamefont {Sherrell}, \citenamefont {Agnoli}, \citenamefont {Craciun},
  \citenamefont {Russo},\ and\ \citenamefont {Mattevi}}]{Reale2017}%
  \BibitemOpen
  \bibfield  {author} {\bibinfo {author} {\bibfnamefont {F.}~\bibnamefont
  {Reale}}, \bibinfo {author} {\bibfnamefont {P.}~\bibnamefont {Palczynski}},
  \bibinfo {author} {\bibfnamefont {I.}~\bibnamefont {Amit}}, \bibinfo {author}
  {\bibfnamefont {G.~F.}\ \bibnamefont {Jones}}, \bibinfo {author}
  {\bibfnamefont {J.~D.}\ \bibnamefont {Mehew}}, \bibinfo {author}
  {\bibfnamefont {A.}~\bibnamefont {Bacon}}, \bibinfo {author} {\bibfnamefont
  {N.}~\bibnamefont {Ni}}, \bibinfo {author} {\bibfnamefont {P.~C.}\
  \bibnamefont {Sherrell}}, \bibinfo {author} {\bibfnamefont {S.}~\bibnamefont
  {Agnoli}}, \bibinfo {author} {\bibfnamefont {M.~F.}\ \bibnamefont {Craciun}},
  \bibinfo {author} {\bibfnamefont {S.}~\bibnamefont {Russo}}, \ and\ \bibinfo
  {author} {\bibfnamefont {C.}~\bibnamefont {Mattevi}},\ }\href {\doibase
  10.1038/s41598-017-14928-2} {\bibfield  {journal} {\bibinfo  {journal}
  {Scientific Reports}\ }\textbf {\bibinfo {volume} {7}},\ \bibinfo {pages}
  {14911} (\bibinfo {year} {2017})}\BibitemShut {NoStop}%
\bibitem [{\citenamefont {Bartolomeo}\ \emph {et~al.}(2017)\citenamefont
  {Bartolomeo}, \citenamefont {Genovese}, \citenamefont {Giubileo},
  \citenamefont {Iemmo}, \citenamefont {Luongo}, \citenamefont {Foller},\ and\
  \citenamefont {Schleberger}}]{DiBartolomeo2017}%
  \BibitemOpen
  \bibfield  {author} {\bibinfo {author} {\bibfnamefont {A.~D.}\ \bibnamefont
  {Bartolomeo}}, \bibinfo {author} {\bibfnamefont {L.}~\bibnamefont
  {Genovese}}, \bibinfo {author} {\bibfnamefont {F.}~\bibnamefont {Giubileo}},
  \bibinfo {author} {\bibfnamefont {L.}~\bibnamefont {Iemmo}}, \bibinfo
  {author} {\bibfnamefont {G.}~\bibnamefont {Luongo}}, \bibinfo {author}
  {\bibfnamefont {T.}~\bibnamefont {Foller}}, \ and\ \bibinfo {author}
  {\bibfnamefont {M.}~\bibnamefont {Schleberger}},\ }\href
  {http://stacks.iop.org/2053-1583/5/i=1/a=015014} {\bibfield  {journal}
  {\bibinfo  {journal} {2D Materials}\ }\textbf {\bibinfo {volume} {5}},\
  \bibinfo {pages} {015014} (\bibinfo {year} {2017})}\BibitemShut {NoStop}%
\end{thebibliography}%
\end{document}